\documentclass[9pt,twocolumn,twoside]{opticajnl}
\journal{opticajournal} 
\usepackage{upgreek}
\setboolean{shortarticle}{true}

\usepackage{multirow}
\usepackage{threeparttable}

\usepackage{lineno}
\usepackage{pdfpages}

\usepackage{booktabs}
\usepackage{multirow}
\usepackage{threeparttable} 
\usepackage{makecell}

\usepackage{enumitem}

\title{Efficient simulation of  millimeter-scale complex-modulated  integrated Bragg gratings via hierarchical locally periodic eigenmode expansion}

\author[1,*]{Rui Cheng}
\author[1]{Jia Meng}
\author[2]{Ping Yu}
\author[1]{Jihao Wang}
\author[1]{Zikun Xie}

\affil[1]{School of Instrument Science and Opto-electronics Engineering, Hefei University of Technology, Hefei, Anhui 230009, China}
\affil[2]{Ningbo University of Technology, Ningbo, 315211, China}

\affil[*]{rcheng@hfut.edu.cn}

\begin{abstract}
We propose a structure-aware, hierarchical locally periodic eigenmode expansion (HLP-EME) framework for efficiently simulating millimeter-scale integrated Bragg gratings (IBGs) with complex modulation on silicon-on-insulator platforms.  
HLP-EME discretizes continuously varying grating parameter profiles into piecewise-constant blocks and exploits the resulting local periodicity of the physical grating structure by reusing the S-matrix of a representative period within each block.
This strategy reduces full-device simulation times for millimeter-scale IBGs to a few minutes, providing a speedup exceeding three orders of magnitude over conventional 3D-FDTD simulations. The method accommodates diverse IBG configurations, including  intra-mode gratings, mode-converting multimode gratings and grating-assisted contra-directional couplers. Experimental results validate the predicted reflection spectra of several complex-modulated IBGs and the reflection phase response of one representative design. The framework is further extended to curved waveguides and successfully captures curvature-induced spectral distortions in millimeter-long, Gaussian-apodized spiral IBGs. Combining full-vectorial modeling with high  computational efficiency, HLP-EME provides a powerful tool for designing and optimizing  long, complex IBG-based photonic devices.
\end{abstract}

\setboolean{displaycopyright}{false} 

\begin{document}
\maketitle

Integrated Bragg gratings (IBGs) on the silicon-on-insulator (SOI) platform are fundamental components in silicon photonics \cite{Chrostowski2015}. Although  uniform gratings  only a few tens of micrometers long can serve as  simple reflective  filters, many applications demand substantially  longer gratings with spatially varying properties. For example, achieving effective sidelobe suppression via apodization \cite{Ma2018, Hung2024}, tailoring complex amplitude and phase responses  for optical signal processing and microwave photonics  \cite{Cheng2018, Kaushal:20}, and realizing sub-nanometer-bandwidth FSR-free filters \cite{Hung2016,Liang2023} all necessitate interaction lengths ranging from hundreds of micrometers to several millimeters. Similarly, chirped gratings for dispersion compensation \cite{Wang2018} and grating-based slow-light waveguides \cite{Xu2024}  rely on millimeter-scale lengths to accumulate sufficient group delay.

Accurate  simulation is therefore critical for designing  such long, non-uniform IBGs. However, 
rigorous structure-aware  methods, such as 3D finite-difference time-domain (FDTD) and standard eigenmode expansion, are computationally prohibitive for millimeter-scale devices. Conversely, fast analytical approaches based on coupled-mode theory (CMT) combined with the transfer matrix method (TMM) cannot capture crucial structure-dependent effects \cite{Cheng21}.
These include the efficacy of geometric modulations for spectral control, unintended geometry-induced index shifts, and  fabrication tolerances.
Moreover, these models are  unable to reliably handle  complex configurations such as mode-converting multimode IBGs \cite{Qiu2016}, grating-assisted contra-directional couplers (CDCs) \cite{Naghdi2017}, and spiral IBGs subject to bend-induced phase distortions \cite{Ma2018}.

To address these challenges, we propose a structure-aware modeling framework based on hierarchical locally periodic eigenmode expansion (HLP-EME). This highly versatile method rapidly simulates millimeter-scale complex-modulated IBGs. It seamlessly accommodates  intra-mode gratings, mode-converting multimode gratings and grating-assisted CDCs, alongside different apodization schemes including lateral phase-delay and corrugation-width modulations. Furthermore, we extend HLP-EME to model spiral IBGs, successfully capturing curvature-induced spectral degradations. 
Experimental results confirm that our framework accurately predicts both the reflection spectra across several complex-modulated IBGs and the phase response of a representative design.
By combining full-vectorial modeling with exceptional computational efficiency, 
HLP-EME facilitates efficient design and optimization of diverse long, complex IBG devices.

\begin{figure*}[t!]
	\centering\includegraphics[width=0.8\linewidth]{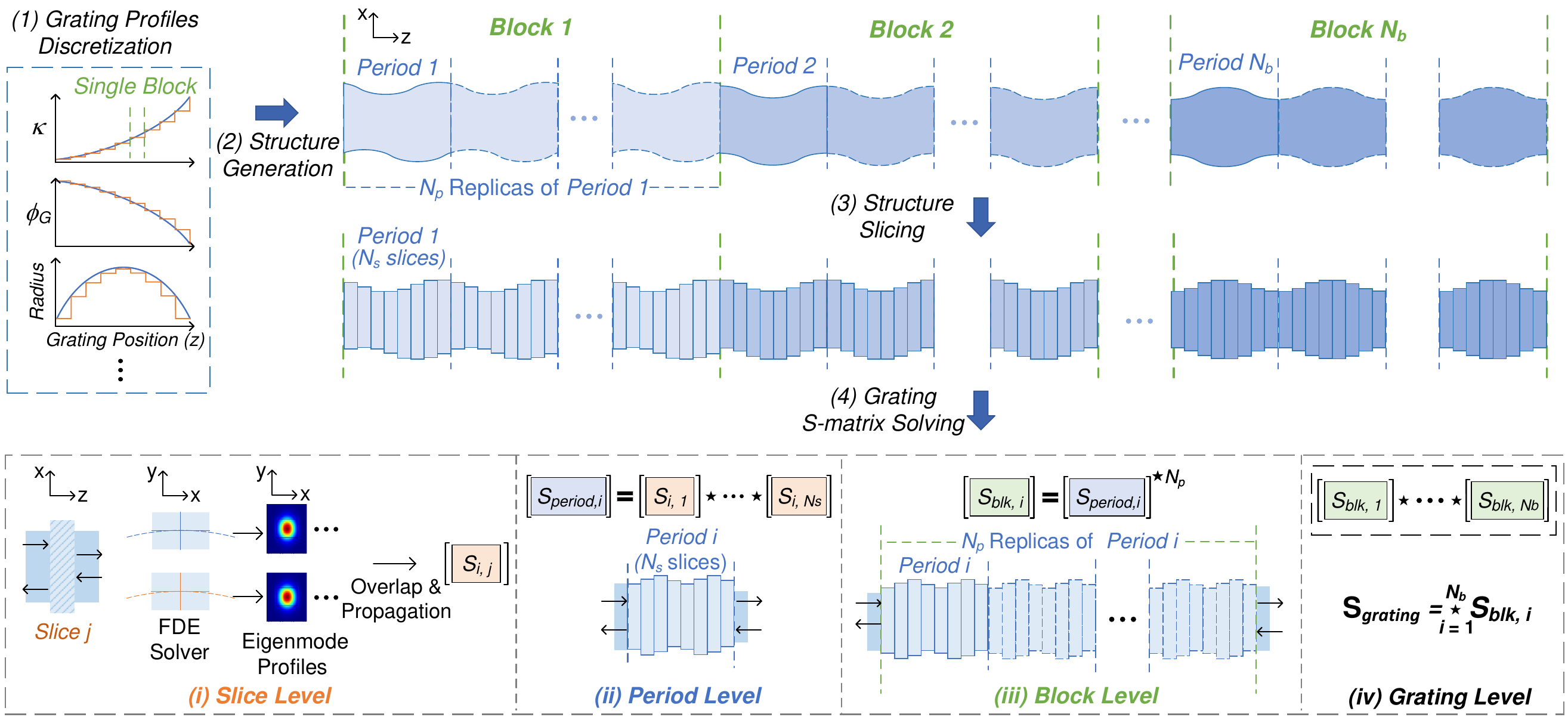}
	\caption{Operating principle of the proposed HLP-EME framework. \textit{Period $i$} denotes the representative grating period of \textit{Block $i$}.}
    \vspace{-0.2cm}
	\label{principle}
\end{figure*}

Figure \ref{principle} illustrates the operating principle of the proposed HLP-EME framework.
The workflow begins with the continuous longitudinal parameter profiles of the grating,  comprising the coupling strength $\kappa(z)$, the grating phase $\phi_G(z)$, and, for spiral gratings, the local bend radius $R(z)$. 
These profiles are synchronously discretized into a sequence of piecewise-constant blocks, each spanning  $N_p$ grating periods.
This parameter-level discretization yields a piecewise-periodic representation of the physical grating structure.
Within each  block, the local parameters are held constant,  so that the block consists of a cascade of $N_p$ identical grating periods.
The resulting local periodicity is the key to the computational efficiency of the proposed method.
For the EME calculation, each representative grating period is further divided along the propagation direction into $N_s$ $z$-invariant waveguide slices. 
The scattering matrix (S-matrix) of the complete device is then assembled through a four-level hierarchical procedure:
\begin{itemize}[itemsep=1pt]
\item \textbf{Slice level:} A finite-difference eigenmode (FDE) solver computes the local mode basis for each slice. Combining the mode-overlap (interface) S-matrix between adjacent slices with the propagation phase of the slice yields the individual slice S-matrix, $S_{i, j}$, which denotes the S-matrix of the $j$-th slice in the representative period of the $i$-th block.
\item \textbf{Period level:}  The $N_s$ slice S-matrices belonging to one grating period are cascaded sequentially to form the  single-period  S-matrix of the $i$-th block, $S_{\mathrm{period},i}$.
\item \textbf{Block level:} 
Since the block is composed of a cascade of $N_p$ identical grating periods, the local eigenmodes and the corresponding  single-period S-matrix ($S_{\mathrm{period},i}$)  are  computed only once for each block. 
The block S-matrix is then obtained by self-cascading $S_{\mathrm{period},i}$:
\begin{equation}
S_{\mathrm{blk},i}=
\underbrace{
S_{\mathrm{period},i}\star S_{\mathrm{period},i}\star\cdots\star S_{\mathrm{period},i}
}_{N_p\ \text{times}}=
S_{\mathrm{period},i}^{\star N_p},
\end{equation}
where $\star$ is the Redheffer star product.   This reuse of the single-period solution substantially reduces the computational cost.
\item \textbf{Grating level:} Finally, cascading the block S-matrices 
$\{S_{\mathrm{blk},i}\}$ 
yields the global S-matrix of the entire IBG, from which the reflection and transmission responses (magnitude, phase, and group delay) are directly extracted.
\end{itemize}

\begin{figure}[ht!]
    \centering
    \includegraphics[width=1\linewidth]{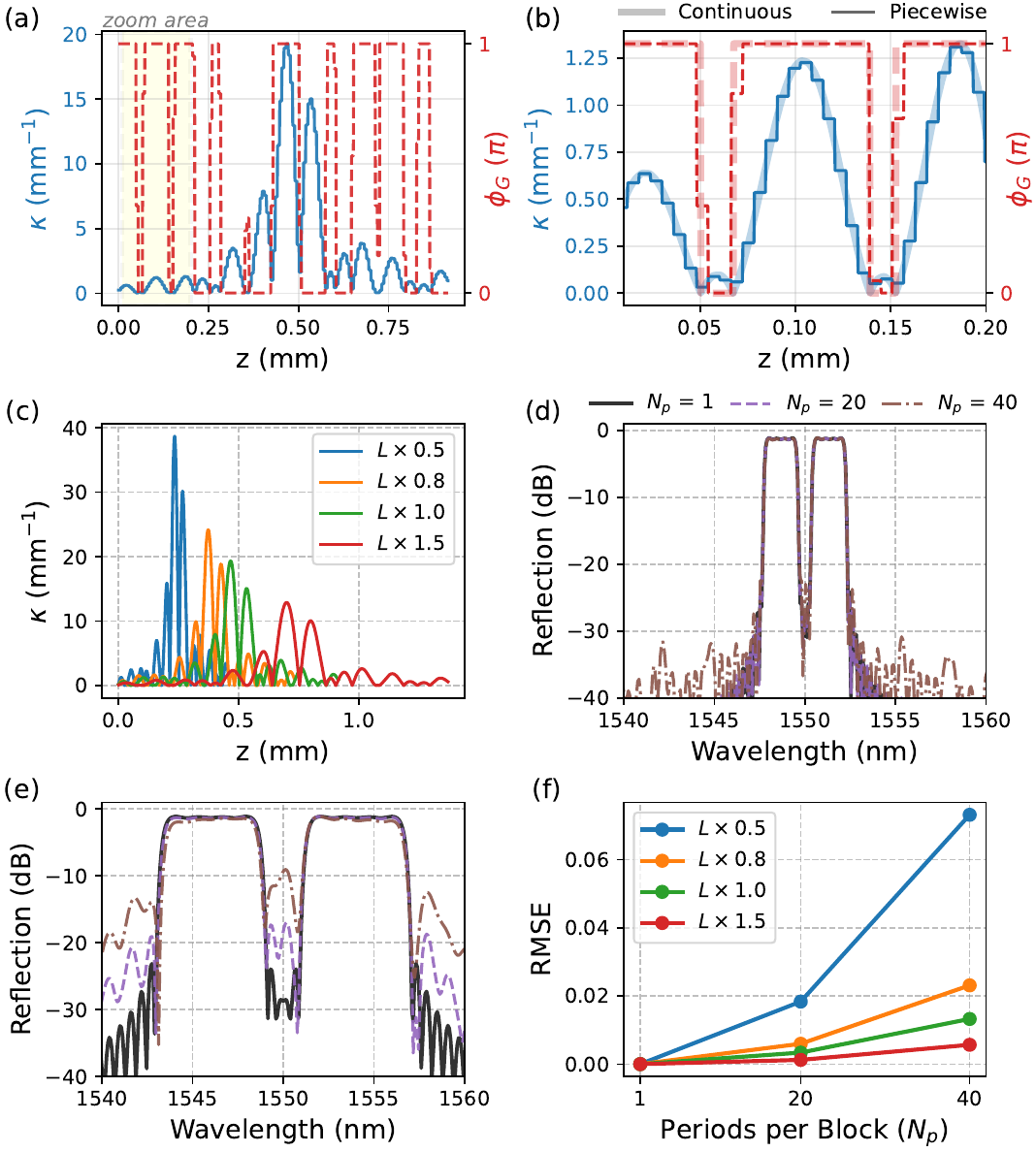}
\caption{(a) Piecewise-constant $\kappa(z)$ and $\phi_G(z)$ at $N_p=20$ (block length $\sim$6 $\upmu$m). (b) Zoom-in of the highlighted region in (a). (c) $\kappa(z)$ profiles with different length factors. Simulated reflection spectra for (d) 1.5 and (e) 0.5 length factors. (f) Spectral RMSE against the $N_p=1$ reference  versus $N_p$ across length factors.}
    \label{np}
\end{figure}

The accuracy and computational efficiency of HLP-EME are primarily governed by two discretization parameters: the block length, $L_{\mathrm{blk}}=N_p\Lambda_G$, where $\Lambda_G$ is the grating period, and  $N_s$. 
We investigate the convergence  over $L_{blk}$ by varying $N_p$  for a two-channel square filter based on a complex-modulated IBG with  $\Lambda_G=302~\mathrm{nm}$. The piecewise-constant  $\kappa(z)$ and  $\phi_G(z)$ profiles for $N_p=20$ are shown in Figs.~\ref{np}(a) and \ref{np}(b). Because the required $L_{\mathrm{blk}}$, or equivalently $N_p$, depends on the longitudinal  variation rate  of the grating profiles, we  scale the original design profiles longitudinally  by different length factors [Fig.~\ref{np}(c)]  to tune this variation rate, with the amplitude of $\kappa(z)$ scaled accordingly. This adjusts the bandwidth accordingly while maintaining the spectral  shape.
The reflection spectra of the piecewise-approximated gratings with different $N_p$ and hence  $L_{\mathrm{blk}}$ are calculated using the CMT-based TMM \cite{Cheng21}. 
Figures~\ref{np}(d) and \ref{np}(e) display the simulated responses for length factors of 1.5 and 0.5, respectively. As shown in Fig.~\ref{np}(f), the root-mean-square error (RMSE) of the spectrum, evaluated against the $N_p=1$ reference, increases with $N_p$. 
Moreover, for a given $N_p$, smaller length factors lead to larger errors due to more rapid longitudinal variations. Nevertheless, $N_p = 20$,   corresponding to  $L_{\mathrm{blk}} \approx 6$ $\upmu$m, consistently ensures high fidelity ($\mathrm{RMSE}<0.02$), offering a good compromise between  accuracy and computational load.

\begin{figure}[t!]
	\centering\includegraphics[width=1\linewidth]{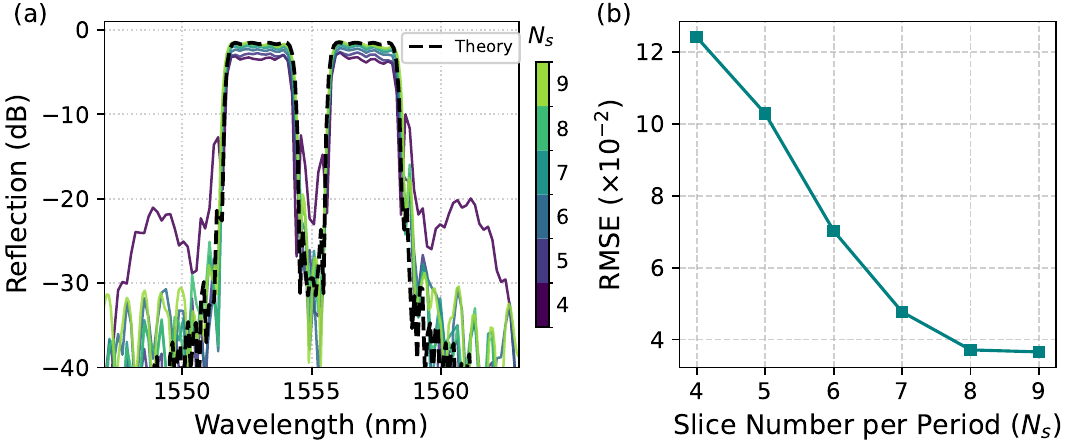}
	\caption{
(a) TE$_0$-TE$_1$ reflection spectra calculated by the HLP-EME method with varying $N_s$, compared with the ideal CMT spectrum. (b) Spectral RMSE between the HLP-EME results and the CMT analytical target as a function of $N_s$.
 }
    \vspace{-0.3cm}
	\label{ns}
\end{figure}

Next, we evaluate the convergence with respect to $N_s$. 
The target grating profiles of the dual-channel square filter are mapped onto a physical asymmetric multimode IBG  with $\mathrm{TE}_0$-to-$\mathrm{TE}_1$ mode conversion via lateral phase-delay modulation (LPDM)~\cite{Cheng21} (parameters detailed in Supplement 1). 
With $N_p=20$ fixed, the reflection spectrum is computed using HLP-EME  for different values of $N_s$. The local modal properties are extracted using a full-vectorial eigenmode solver (Lumerical MODE), although  the HLP-EME formulation   is platform-independent.
The HLP-EME   spectra  converge toward the CMT analytical target as $N_s$ increases [Fig.~\ref{ns}(a)]. The corresponding spectral RMSE decreases rapidly and saturates for $N_s\geq8$ [Fig.~\ref{ns}(b)], indicating numerical convergence.  
Consequently, adopting $L_{\mathrm{blk}} \approx 6~\upmu\mathrm{m}$ and $N_s = 8$ minimizes computational overhead while preserving high spectral fidelity. Since $\Lambda_G \approx 300~\mathrm{nm}$ ($292$--$332~\mathrm{nm}$) in this work, $N_p = 20$ and $N_s = 8$ are used throughout  subsequent simulations unless stated otherwise.

\begin{figure}[t!]
	\centering\includegraphics[width=1\linewidth]{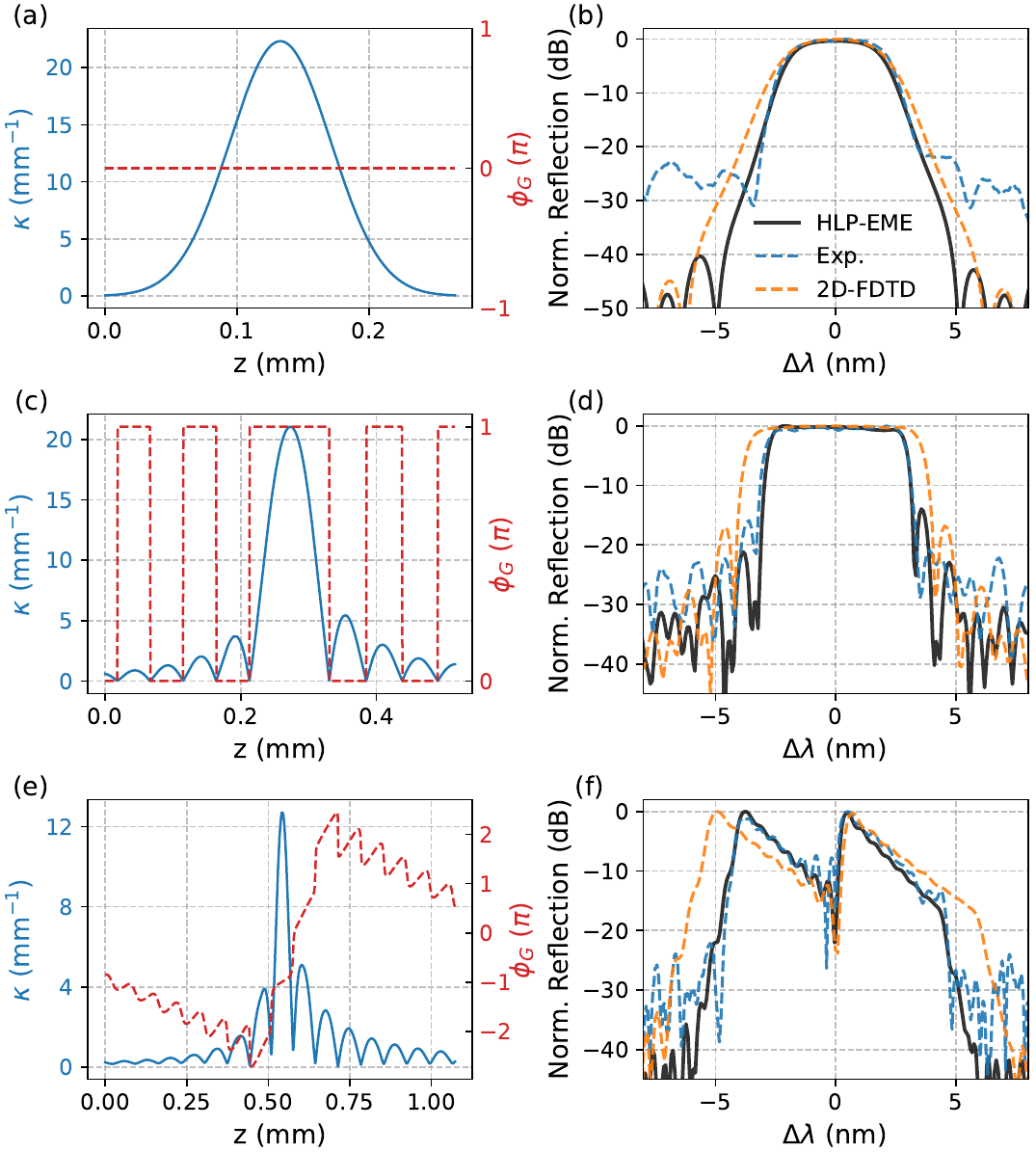}
	\caption{
		Comparison of simulated and experimental results for various complex-modulated multimode IBGs. Grating parameter profiles (left panels) and the corresponding normalized TE$_0$-TE$_1$ reflection spectra (right panels) for: (a, b) Gaussian-apodized grating, (c, d) single-channel square filter, and (e, f) dual-channel linear edge filter.
	}
	\label{exp1}
	    \vspace{-0.2cm}

\end{figure}
To validate the predictive accuracy of the proposed HLP-EME, various complex-modulated IBGs were designed, fabricated, and experimentally characterized (see Supplement 1 for detailed device parameters, fabrication, and experimental setup). 
We first examine three asymmetric multimode  IBGs featuring $\text{TE}_0$-to-$\text{TE}_1$ mode conversion, where apodization is implemented through LPDM. These devices comprise a Gaussian-apodized sidelobe-suppressed filter, a single-channel square filter, and a dual-channel linear edge filter. Their respective grating profiles are depicted in the left panels of Fig.~\ref{exp1}. As shown in the corresponding right panels, the simulated spectra from the HLP-EME framework exhibit excellent agreement with the experimental measurements, verifying its reliability in modeling long, highly complex IBGs. In contrast, while the 2D-FDTD simulations preserve the overall spectral profile, they yield noticeably broader reflection bandwidths, which is attributed to an overestimation of the mode coupling strength.

\begin{table}[ht!]
\centering
\caption{Simulation Time and Speedup Comparison}
\label{tab:performance_comparison}
\begin{threeparttable}
\setlength{\tabcolsep}{6pt}
\begin{tabular}{@{}lccc@{}}
\toprule
\multirow{2}{*}{\textbf{Type ($L$ in mm)}} & \multicolumn{3}{c}{\textbf{Runtime (Speedup)}} \\
\cmidrule(l){2-4}
& \textbf{HLP-EME} & \textbf{2D-FDTD} & \textbf{3D-FDTD\tnote{*}} \\
\midrule
Gau.  (0.266) & \textbf{28 s} & 4.6 min ($10\times$) & 16.8 h ($2160\times$) \\
 Square (0.516) & \textbf{1.2 min} & 24 min ($20\times$) & 75.8 h ($3790\times$) \\
Edge (1.072) & \textbf{5.9 min} & 2.2 h ($22\times$) & 360 h ($3661\times$) \\
\bottomrule
\end{tabular}
\begin{tablenotes}
\item[*] \footnotesize 3D-FDTD times are extrapolated ($O(L^2)$) due to extreme lengths.
\end{tablenotes}
\end{threeparttable}
    \vspace{-0.2cm}
\end{table}

The computational costs for these three multimode  IBGs are summarized in Table~\ref{tab:performance_comparison}. Because directly simulating such extreme lengths using 3D-FDTD is impractical, the baseline 3D-FDTD times were extrapolated. Simulation setting, hardware and  extrapolation details are provided in  Supplement 1. The proposed HLP-EME framework delivers dramatic efficiency improvements relative to 3D-FDTD: simulation times are reduced from hundreds of hours down to  minutes, yielding speedups exceeding three orders of magnitude ($>2160\times$). 
Furthermore, HLP-EME is roughly an order of magnitude faster than 2D-FDTD. Importantly, unlike 2D-FDTD, which relies on dimensional reduction, the HLP-EME method performs full 3D simulations without dimensional approximations.

\begin{figure}[t!]
	\centering\includegraphics[width=1\linewidth]{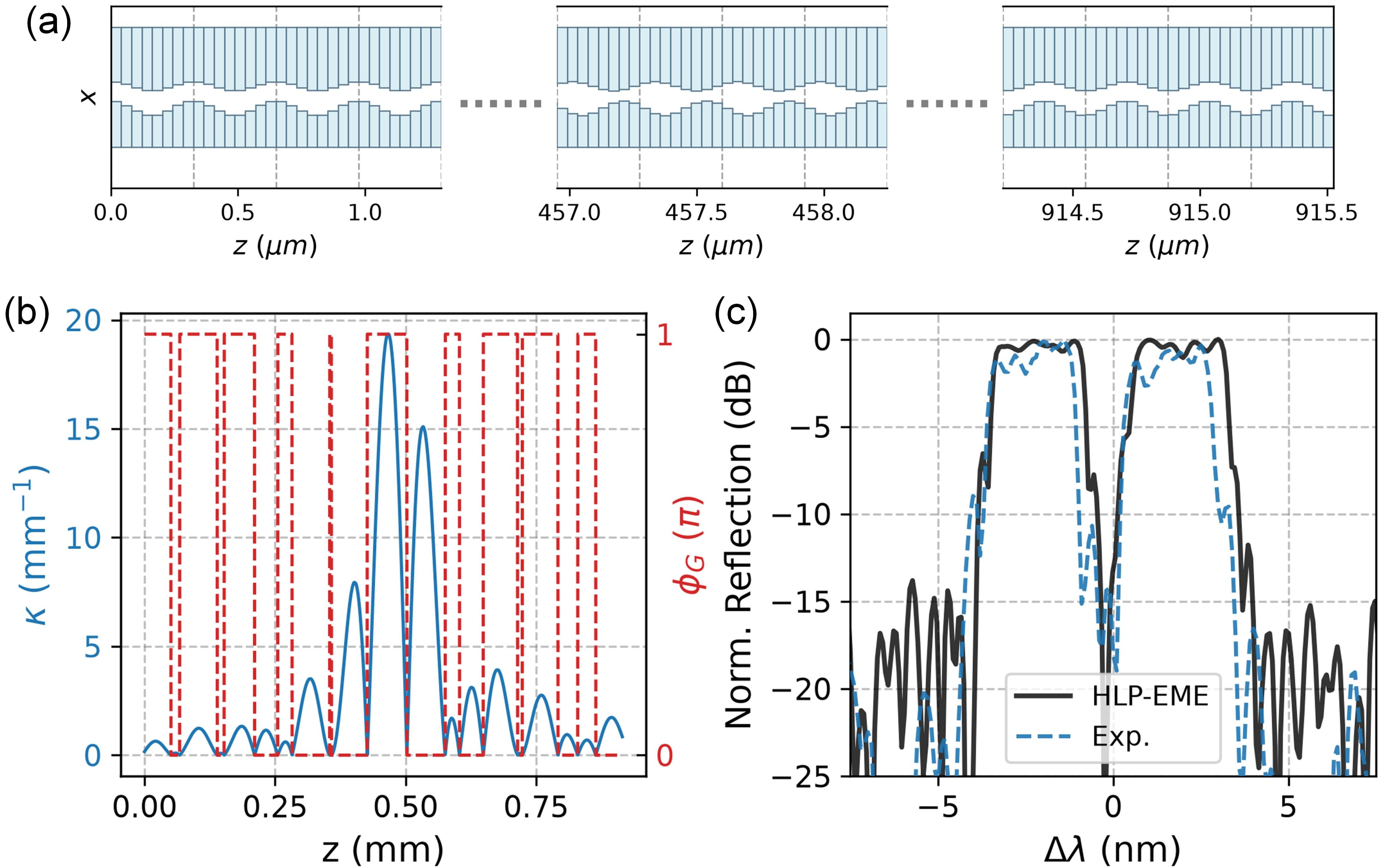}
\caption{(a) Schematic of the sliced grating-assisted CDC structure. 
The corrugation widths are exaggerated by a factor of five for clarity. (b) $\kappa(z)$ and $\phi_G(z)$ profiles. (c) Measured versus HLP-EME-simulated normalized reflection spectra.}
	\label{cdc}
\end{figure}

To demonstrate the versatility of the proposed HLP-EME, we extend the method to model grating-assisted CDCs. The simulation follows a similar longitudinal slicing procedure, but is applied across the composite cross-section of a coupled dual-grating system [Fig.~\ref{cdc}(a)]. The device operates as a dual-channel square filter [Fig.~\ref{cdc}(b)] and is apodized via LPDM. 
As shown in Fig.~\ref{cdc}(c), the HLP-EME-simulated spectrum agrees well with the experimental measurement.

\begin{figure}[t!]
	\centering\includegraphics[width=1\linewidth]{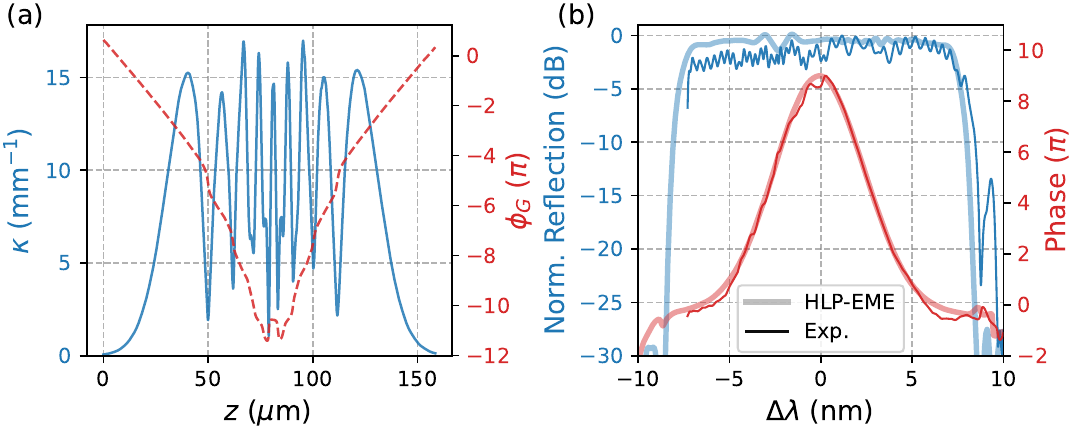}
\caption{(a) $\kappa(z)$ and $\phi_G(z)$ profiles for an IBG square filter with a Gaussian phase response. (b) Measured and HLP-EME-simulated normalized reflection power and phase responses.}
	\label{gau_phase}
	    \vspace{-0.2cm}
\end{figure}

Beyond amplitude spectra prediction, the proposed method rigorously simulates phase responses of  complex IBGs. To verify this, we model a  square filter with a customized Gaussian-shaped phase response [Fig.~\ref{gau_phase}(a)]. This filter is developed on an  intra-mode ($\text{TE}_0$-to-$\text{TE}_0$) IBG apodized using LPDM. 
The measured and HLP-EME-simulated spectra are compared in Fig.~\ref{gau_phase}(b), with the phase responses offset to the same baseline for visual comparison. Both the power and phase profiles demonstrate high consistency.

\begin{figure}[t!]
    \centering\includegraphics[width=0.97\linewidth]{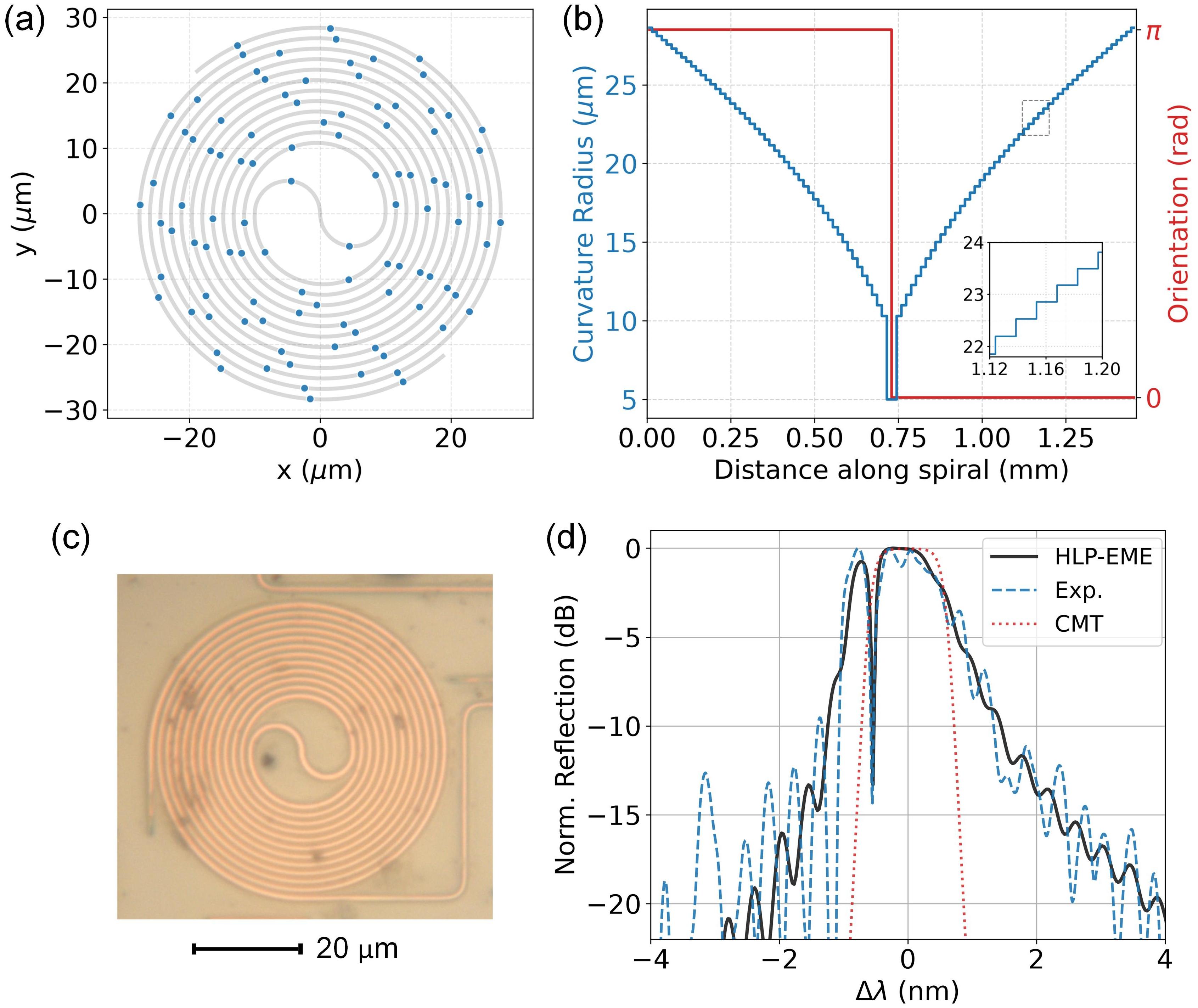}
\caption{(a) Analytical spiral path of the grating; blue dots indicate centers of piecewise-constant blocks. (b) Piecewise-uniform curvature radius and orientation profiles. (c) Optical micrograph of the fabricated spiral IBG. (d) Measured and HLP-EME-simulated normalized reflection spectra, alongside the CMT-based TMM ideal target.}
    \label{spiral}
		    \vspace{-0.2cm}

\end{figure}

Finally, we deploy HLP-EME to simulate a millimeter-long, LPDM Gaussian-apodized intra-mode ($\text{TE}_0$-to-$\text{TE}_0$) spiral IBG [Fig.~\ref{spiral}(a) and \ref{spiral}(c)]. 
The continuous local radius $R(z)$, bending orientation  and $\kappa(z)$  are discretized into piecewise-constant blocks  [Figs.~\ref{spiral}(b)], with $N_p = 50$  chosen due to the smaller longitudinal gradients of these parameters.
For each slice, an FDE solver formulated in cylindrical coordinates computes the modes at the local bend radius. The measured reflection spectrum closely matches the HLP-EME prediction  [Fig. \ref{spiral}(d)]. Notably, both exhibit an identical departure from the ideal target computed via CMT-based TMM, which stems from bend-induced perturbations in the local grating parameters. 
These results confirm that HLP-EME accurately captures such distortions, providing an effective framework for characterizing and pre-compensating bend-induced grating spectral degradations.

In conclusion, we have demonstrated a  versatile, structure-aware HLP-EME framework that accelerates the simulation of millimeter-scale complex-modulated IBGs by more than three orders of magnitude over 3D-FDTD, without compromising full-vectorial accuracy. Experimental results validate its predictions  across diverse configurations, including intra-mode gratings, asymmetric multimode gratings with mode conversion, and grating-assisted CDCs. The framework also reliably reproduces curvature-induced spectral distortions in millimeter-long, apodized spiral IBGs, establishing HLP-EME as an efficient, high-fidelity design tool for long, complex IBG devices.
Future work will quantify the relationship between the grating profile gradient and the required discretization resolution to guide adaptive, nonuniform  discretization with variable block lengths.

\smallskip \noindent\textbf{Funding.}
\small{National Natural Science Foundation of China (62105089).}

\smallskip \noindent\textbf{Disclosures.} \small{The authors declare no conflicts of interest.}

\smallskip \noindent\textbf{Code availability.} \small{The source code of the  model is publicly available  at \url{https://github.com/ruicheng-photonics/hlp-eme}.}

\smallskip \noindent\textbf{Data availability.} Data underlying the results presented in this paper are available from the authors upon reasonable request.

\smallskip \noindent\textbf{Supplemental document.} \small{See Supplement 1 for supporting content.}

\bibliography{refs}




\end{document}